%% file: main.tex
\documentclass[fleqn,11pt]{article}

\usepackage{amsfonts,amsmath,amssymb}
\usepackage{latexsym}
\usepackage[square,numbers,sort&compress]{natbib}
\usepackage{orcidlink}
\usepackage{commath}
\usepackage{float}
\hypersetup{colorlinks,citecolor=blue}
\hypersetup{colorlinks=true,linkcolor=red,filecolor=magenta, urlcolor=blue}
\usepackage{mathtools,amssymb,lipsum}
\usepackage{cuted}
\usepackage{etoolbox}
\input preamble.tex

\begin{document}
\twocolumn[



\Title{Late-Time Cosmic Acceleration in Ricci–Gauss–Bonnet Gravity via Gradient Descent Optimization}

\Aunames{Santosh V. Lohakare\,\orcidlink{0000-0001-5934-3428}\au{a,1} , B. Mishra\,\orcidlink{0000-0001-5527-3565}\au{b,2}, S. K. Maurya\,\orcidlink{0000-0003-4089-3651}\au{c,d,3}}

\Addresses{
\adr a {Department of Mathematics, School of Advanced Sciences, Vellore Institute of Technology, Vellore, 632014, Tamilnadu, India}
\adr b {Department of Mathematics,
Birla Institute of Technology and Science-Pilani, Hyderabad Campus, Jawahar Nagar, Kapra Mandal, Medchal District, Telangana 500078, India}
\adr c {Department of Mathematical and Physical Sciences, College of Arts and Sciences, University of Nizwa, P.O. Box 33, Nizwa 616, Sultanate of Oman}
\adr d {Research Center of Astrophysics and Cosmology, Khazar University, Baku, AZ1096, 41 Mehseti Street, Azerbaijan}
}


\Abstract
	{\textbf{Abstract:} We study the late-time evolution of the Universe within the $f(R,\mathcal{G})$ gravity framework, where $R$ is the Ricci scalar and $\mathcal{G}$ is the Gauss–Bonnet term. To make the model tractable, we propose a parametrization scheme and determine its parameters using Gradient Descent, with constraints coming from the latest Cosmic Chronometer (CC) and Pantheon$^+$ supernovae data. Key cosmological indicators—the deceleration parameter $(q)$ and the equation of state $(\omega)$—show a clear shift from past deceleration to the present accelerated expansion. Interestingly, the EoS parameter $\omega$ remains above the phantom divide, indicating quintessence-like behavior in agreement with current observations. Energy condition assessments reinforce this framework: the strong energy condition is violated, which is consistent with models permitting cosmic acceleration, whereas both the weak and null energy conditions hold true. To check consistency, we also apply the $\mathrm{Om}(z)$ diagnostic, which separates this model from the simple cosmological constant case and highlights its favoring of a quintessence-dominated future. Using the best-fit values, we estimate the age of the Universe, which comes out in close agreement with independent astrophysical measurements. Taken together, the results suggest that $f(R,\mathcal{G})$ gravity provides a credible and self-consistent route to explain late-time cosmic acceleration, especially when tested against the combined CC and Pantheon$^+$ datasets.  
	
    }
\medskip

]
\email 1 {lohakaresv@gmail.com}
\email 2 {bivu@hyderabad.bits-pilani.ac.in}
\email 3 {sunil@unizwa.edu.om}

\section{Introduction} \label{SEC-I}
    The mystery of why the Universe is expanding at an accelerating rate remains one of the deepest challenges in modern cosmology, urging us to rethink our understanding of gravity and the cosmic timeline. In 1998, observations of Type Ia supernovae, which act as cosmic distance markers due to their consistent brightness, revealed this unexpected speedup \cite{Riess_1998_116, Perlmutter_1998_517}. Recent advances in observational cosmology—including results from the Dark Energy Spectroscopic Instrument (DESI) surveys \cite{Adame_2024_DESI_collaboration, Plaza_2025_112, Bansal_2025_112, Chaudhary_2026_50, Duchaniya_2025_12_jcap}, Type Ia supernovae samples \cite{Riess_1998_116, Perlmutter_1998_517}, measurements from the Wilkinson Microwave Anisotropy Probe (WMAP) \cite{Spergel_2003_148}, and detailed mapping of the cosmic microwave background (CMB) \cite{Hinshaw_2013_208}—have robustly confirmed the late-time accelerated expansion of the Universe. Additional data from large-scale structure probes such as the Baryon Oscillation Spectroscopic Survey (BOSS) \cite{Alam_2017_470} and Baryon Acoustic Oscillation (BAO) measurements \cite{Eisenstein_2005_633} strengthen the case for an explanation beyond the standard framework of General Relativity (GR). These empirical developments have motivated the investigation of numerous modified gravity models, including $f(R)$ gravity \cite{Nojiri_2007_04, Sotiriou_2010_82}, in which the Lagrangian is generalized to an arbitrary function of the Ricci scalar $R$; $f(T)$ gravity \cite{Ferraro_2007_75}, which modifies the underlying teleparallel geometry by considering a function of the torsion scalar $T$; and more recently, $f(Q)$ theories \cite{Jimenez_2018_98, Heisenberg_2023_1066_Review}, based on nonmetricity scalar $Q$. Each of these frameworks attempts to address limitations in standard cosmology and provide a better fit to the wealth of new observational data.
    
    One possibility is an exotic form of matter with unusual properties, like negative pressure combined with positive energy density, acting as a repulsive force. Alternatively, modifying the geometric framework of GR may provide an explanation for the observed acceleration, leading to the development of alternative theories of gravity. The simplest model, $\Lambda$CDM (Lambda Cold Dark Matter), uses a cosmological constant ($\Lambda$) alongside cold dark matter to describe this phenomenon. It aligns well with observations, from the rotation curves of spiral galaxies \cite{Baudis_2016_43} to the early inflationary expansion of the Universe \cite{Perenon_2015_11}. However, $\Lambda$CDM faces challenges, such as the fine-tuning problem, where the cosmological constant’s tiny value seems improbably precise \cite{Carroll_2001_4, weinberg_2008}. This raises questions about whether $\Lambda$ genuinely serves as the main factor or is just a useful substitute.

    One promising approach to go beyond Einstein’s theory is to modify the geometric sector of the field equations rather than introducing additional matter fields. A well-known example is $f(R)$ gravity, where the Einstein–Hilbert action is generalized to an arbitrary function of the Ricci scalar $R$ \cite{Carroll_2004_70, Nojiri_2011_505}. GR is recovered for $f(R)=R$, but more general choices of $f(R)$ yield richer dynamics that can naturally accommodate both the early inflationary phase and the present accelerated expansion. A further extension, known as $f(R,\mathcal{G})$ gravity, incorporates the Gauss–Bonnet invariant $\mathcal{G} \equiv R^2 - 4R^{\mu\nu}R_{\mu\nu} + R^{\mu\nu\sigma\rho}R_{\mu\nu\sigma\rho},$ which allows the action to probe higher-order curvature corrections \cite{Nojiri_2005_631, Cognola_2006_73, Lohakare_2024_MNRAS}. The inclusion of $\mathcal{G}$ is particularly well motivated, as Starobinsky’s pioneering work \cite{Starobinsky_1980_91} showed that curvature-squared terms can successfully describe inflation, while subsequent studies have demonstrated that suitable $f(R,\mathcal{G})$ couplings may also explain the late-time acceleration associated with dark energy \cite{Benetti_2018_27, Martino_2020_102}. Nevertheless, these theories must be carefully examined for stability and for consistency with cosmological data, including the latest Planck results \cite{Capozziello_2014_29}.
    
    Scalar-tensor gravity enriches the framework of gravitational theories by integrating the fourth-order contributions arising from the Gauss-Bonnet invariant with the second-order dynamics introduced by a scalar field. This synthesis establishes a more complex phase space, enhancing our ability to analyze and understand the nuances of cosmic evolution in various gravitational contexts \cite{Konstantinos_2022_2211.06076}. Dynamical system analyses, such as those by Shah et al. \cite{Shah_2019_79}, reveal how these models handle transitions between decelerating and accelerating phases. Some $f(R)$ gravity models, like the Starobinsky model with $n=2$, fit observational data—such as the Hubble parameter $H(z)$ and growth rate $[f\sigma_8] (z)$—as well as or better than $\Lambda$CDM \cite{Bessa_2022_82}. Regularized Einstein-Gauss-Bonnet gravity, explored by Bayarsaikhan et al. \cite{Bayarsaikhan_2023_83}, further refines this with non-minimal scalar coupling, showing promise but raising questions about the absence of a linear Ricci scalar term in some formulations.

    In this work, we study a specific $f(R,\mathcal{G})$ model of the form $f(R,\mathcal{G}) = R + \alpha R^2 + \beta \mathcal{G}^2$, where $\alpha$ and $\beta$ are free parameters to be constrained \cite{Martino_2020_102}. To make the model tractable, we introduce a parametrization scheme and determine its parameters using the Gradient Descent method applied to the latest observational datasets. In particular, we use Cosmic Chronometer (CC) measurements of the Hubble parameter \cite{Moresco_2022_25} and the Pantheon$^+$ supernovae catalogue \cite{Brout_2022_938}, which together provide tight bounds on the expansion history. Section \ref{SEC-II} presents the theoretical framework of $f(R,\mathcal{G})$ gravity, while Section \ref{SEC-3} details the fitting procedure and the resulting parameter constraints. In Section \ref{SEC-4}, we analyze the cosmological behavior of the model through the deceleration parameter $q$ and the effective equation of state $\omega$, which show the expected transition from past deceleration to present acceleration without crossing the phantom divide. We further test stability and consistency through energy condition analyses, finding that the weak and null conditions hold while the strong condition is violated, as expected in accelerating scenarios in Section \ref{energy conditions}. Additional checks with the $\mathrm{Om}(z)$ diagnostic distinguish our model from $\Lambda$CDM and highlight its quintessence-like future evolution, while age estimates of the Universe from the best-fit parameters agree closely with independent astrophysical measurements. Finally, Section \ref{SEC-5} summarizes these results and discusses their broader implications for understanding late-time cosmic acceleration.

\section{Mathematical Formalism of \texorpdfstring{$f(R, \mathcal{G})$}{} Gravity} \label{SEC-II}
    The action considered in $f(R, \mathcal{G})$ gravity represents an extension of GR, inspired by attempts to better capture late-time cosmic acceleration and other phenomena \cite{Laurentis_2015_91, Wu_2015_92, Santos_da_Costa_2018_35, ODINTSOV_2019_938_935, Kumar_Sanyal_2020_37, Brout_2022_938, Lohakare_2023_40_CQG}. This action takes the form
        \begin{equation}\label{1}
            S = \int d^4x\, \sqrt{-g} \left[ \frac{1}{2\kappa^2} f(R,\mathcal{G}) + \mathcal{L}_\mathrm{m} \right],
        \end{equation}
    where $g$ denotes the determinant of the metric tensor, $\mathcal{L}_\mathrm{m}$ is the Lagrangian for matter fields, $\kappa^2=8\pi G$, and $G$ is gravitational constant.

    The Gauss-Bonnet curvature term $\mathcal{G}$ is defined as
        \begin{equation}\label{Eq: Gauss-Bonnet}
            \mathcal{G} \equiv R^2-4R^{\mu \nu} R_{\mu \nu}+R^{\mu \nu \sigma \rho}R_{\mu \nu \sigma \rho} \, ,
        \end{equation}
    where $R^{\mu \nu}$ and $R^{\mu \nu \sigma \rho}$ represent the Ricci and Riemann curvature tensors, respectively.

    In the language of differential geometry, the Gauss-Bonnet invariant possesses a special property: when integrated over a compact four-dimensional manifold $\mathcal{M}$, it is related to the Euler characteristic $\chi(\mathcal{M})$, a topological quantity, as
        \begin{equation}\label{3}
            \int_{\mathcal{M}} \mathcal{G}\, d^{n}x = \chi(\mathcal{M}).
        \end{equation}
    In four dimensions, the Gauss-Bonnet term acts as a topological surface term; including it linearly in the action does not influence the field equations due to its independence from the metric. As a result, any meaningful dynamical contribution from $\mathcal{G}$ requires either a non-linear dependence or a coupling with other curvature invariants or fields \cite{Laurentis_2015_91, Santos_da_Costa_2018_35}.

    The field equations for $f(R,\mathcal{G})$ gravity become significantly more complex than Einstein's equations due to higher-order derivative terms. The complete expression involves multiple geometric quantities. By taking the variation of the action in Eq.~(\ref{1}) with respect to the metric tensor $g_{\mu \nu}$, the resulting field equations for $f(R,\mathcal{G})$ gravity are obtained as
        \begin{eqnarray} \label{Eq: 2}
            \hspace{-1cm}& & f_R{G}_{\mu\nu} = \kappa^2 T_{\mu\nu}+\frac{1}{2}g_{\mu\nu}[f(R,\mathcal{G})-Rf_{R}]+\nabla_{\mu}\nabla_{\nu} f_{R} \nonumber \\ \hspace{-1cm}& & -g_{\mu\nu} \Box f_{R}  +2(\nabla_{\mu}\nabla_{\nu}f_\mathcal{G})R -4(\nabla_{k} \nabla_{\mu} f_\mathcal{G})R^{k}_{\nu} \nonumber \\ \hspace{-1cm}& & - 2g_{\mu \nu}(\Box f_\mathcal{G})R + 4(\Box f_\mathcal{G})R_{\mu\nu} + f_\mathcal{G}\Big({-2R}{R_{\mu\nu}} \nonumber \\ \hspace{-1cm}& & + 4 R_{\mu k}R^{k}_{\nu} - 2R^{klm}_{\mu}R_{\nu k l m} + 4 g^{kl} g^{mn} R_{\mu k \nu m} R_{ln}\Big) \nonumber \\ \hspace{-1cm}& & - 4(\nabla_{k} \nabla_{\nu} f_\mathcal{G})R^{k}_{\mu} -4(\nabla_{l} \nabla_{n} f_\mathcal{G})g^{kl}g^{mn}R_{\mu k \nu m} \nonumber \\ \hspace{-1cm}& &+4g_{\mu \nu}(\nabla_{k} \nabla_{l} f_\mathcal{G})R^{kl},
        \end{eqnarray}

    Here, $G_{\mu \nu}$ denotes the Einstein tensor, $\nabla_{i}$ is the covariant derivative compatible with the metric $g_{\mu \nu}$, and $\Box \equiv g^{\mu \nu} \nabla_{\mu} \nabla_{\nu}$ defines the covariant d'Alembert operator. These equations contain fourth-order derivatives of the metric, making $f(R,\mathcal{G})$ gravity theories inherently more complex than standard GR. The terms involving $f_\mathcal{G}$ represent contributions from the Gauss-Bonnet modifications. The term $T_{\mu \nu}$ corresponds to the energy-momentum tensor of matter fields. Furthermore, the partial derivatives of the function $f(R, \mathcal{G})$ with respect to its arguments are defined as follows
        \begin{equation*}
            f_R \equiv \frac{\partial f(R, \mathcal{G})}{\partial R}, \hspace{1cm} f_\mathcal{G} \equiv \frac{\partial f(R, \mathcal{G})}{\partial \mathcal{G}}.
        \end{equation*}

    The background spacetime is described by the spatially flat Friedmann–Lemaître–Robertson–Walker (FLRW) metric, given by  
        \begin{equation} \label{Eq: flrw}
            ds^{2} = -dt^{2} + a^{2}(t)\left(dx^{2} + dy^{2} + dz^{2}\right),
        \end{equation}
    where $a(t)$ is the scale factor, and the Hubble parameter is defined as $H \equiv \frac{\dot{a}(t)}{a(t)}$. Here, an over-dot denotes differentiation with respect to cosmic time $t$.  

    In the FLRW background, both $R$ and $\mathcal{G}$ become functions solely of the Hubble parameter $H$ and its time derivative $\dot{H}$. This reduction allows us to express the entire cosmic evolution in terms of $H(z)$, making observational comparison straightforward. In this background, the Ricci scalar and the Gauss–Bonnet invariant respectively takes the following form  
        \begin{equation} \label{eq: R and G for flrw}
            R = 6(\dot{H} + 2H^{2}), \hspace{1cm} \mathcal{G} = 24H^{2}(\dot{H} + H^{2}).
        \end{equation}

    To model the matter content, we consider an isotropic and homogeneous perfect fluid, whose energy-momentum tensor is given by  
        \begin{equation} \label{7}
            T^{\nu}_{\mu} = \mathrm{diag}(-\rho, p, p, p),
        \end{equation}
    where $\rho$ is the energy density and $p$ denotes the isotropic pressure of the fluid.

    Substituting the metric \eqref{Eq: flrw} and the curvature scalars from \eqref{eq: R and G for flrw} into the modified gravitational field Eq.~\eqref{Eq: 2}, the resulting field equations for $f(R,\mathcal{G})$ gravity in the flat FLRW background can be obtained as
        \begin{eqnarray} \label{4}
            && 3H^{2}f_{R}=\kappa \rho+\frac{1}{2}\big[Rf_{R}+\mathcal{G} f_\mathcal{G}- f(R,\mathcal{G})\big] \nonumber\\ && - 12H^{3}\dot{f}_{\mathcal{G}} - 3H \dot{f}_{{R}},\\ \label{5}
            && 2\dot{H}f_{R}+3H^{2}f_{R}=-\kappa p+\frac{1}{2}\big[Rf_{R}+\mathcal{G} f_\mathcal{G}\nonumber\\ && -f(R,\mathcal{G})\big] -8H\dot{H}\dot{f}_{\mathcal{G}} -2H\dot{f}_{R} -\ddot{f}_{{R}} \nonumber\\ &&- 8H^{3}\dot{f}_{\mathcal{G}} - 4H^{2}\ddot{f}_{\mathcal{G}}
        \end{eqnarray}

    The Einstein field equations, when applied to the FLRW metric \eqref{Eq: flrw}, yield the field equations in the following form
    \begin{eqnarray}
            3 H^2 &=&\kappa^2 \left(\rho_{\text{m}}+ \rho_{\text{DE}}\right) = \kappa^2 \rho_{\text{eff}},  \label{first_field_equation}\\
            \left(2 \dot{H}+3 H^2\right)& =&-\kappa^2 \left(p_{\text{DE}}\right) = -\kappa^2 p_{\text{eff}}, \label{second_field_equation}
    \end{eqnarray}
    where $\rho_{\rm m}$ and $\rho_{\rm DE}$ denote the matter density and the dark energy density, respectively.

    The energy density and matter pressure can be obtained if the functional, $f(R,\mathcal{G})$ has some explicit form. In this study, we consider a class of modified gravity models in which the Ricci scalar $R$ couples additively to a non-linear function of the Gauss-Bonnet invariant $\mathcal{G}$. This structure ensures that the deviation from GR has a genuine impact on the dynamics of the Universe. Specifically, we adopt a separable functional form $f(R, \mathcal{G}) = f_1(R) + f_2(\mathcal{G})$, where both $f_1$ and $f_2$ are chosen to be quadratic functions \cite{Laurentis_2015_91, Lohakare_2023_39}. The linear term in $f_1(R)$ is retained to recover the correct weak-field limit and to ensure consistency with general relativistic behavior at low curvatures.

    We focus on a model that extends the well-known Starobinsky-type correction $R^2$ by including a leading-order non-linear Gauss-Bonnet contribution. Since the linear term in $\mathcal{G}$ does not contribute to the four-dimensional field equations due to its topological nature, the first dynamically significant correction arises from $\mathcal{G}^2$. This leads to the following specific choice of the functional form
        \begin{equation} \label{6}
            f(R, \mathcal{G}) = R + \alpha R^2 + \beta \mathcal{G}^{2},
        \end{equation}
    where $\alpha$ and $\beta$ are free constants that determine the strength of the respective curvature corrections \cite{Martino_2020_102, Lohakare_2021_96, Laurentis_2015_91}. Substituting Eq.~\eqref{6} into Eqs.~\eqref{4} and \eqref{5}, the corresponding expressions for the energy density and pressure in terms of the Hubble parameter can be derived as follows
        \begin{eqnarray} 
            \hspace{-3.5cm}\rho &=& \frac{1}{\kappa^{2}}(3H^2+108\alpha \dot{H} H^2+1728\beta \dot{H} H^6 \nonumber \\ \hspace{-3.5cm} & & + 864\beta \dot{H^2}H^4+36\alpha H\ddot{H}+576\beta\ddot{H}H^5 \nonumber\\ & &-18\alpha \dot{H^2}-288\beta H^8), \label{density field eq} \\
            \hspace{-3.5cm} p &=& \frac{1}{\kappa^{2}}(-2\dot{H}-3H^2-54\alpha \dot{H^2}-108\alpha \dot{H} H^2 \nonumber \\ \hspace{-3.5cm} & & - 960\beta \dot{H} H^6-4320\beta \dot{H^2}H^4-72\alpha H\ddot{H}\nonumber \\ \hspace{-3.5cm} & & - 1152\beta \ddot{H}H^5-1152\beta H^2 \dot{H^3} -12\alpha \dot{\ddot{H}} \nonumber\\ \hspace{-3.5cm} & & - 192\beta \dot{\ddot{H}}H^4 + 288\beta H^8- 1536\beta \dot{H} \ddot{H} H^3) \, .\nonumber\\ \label{pressure field eq}
        \end{eqnarray}

    The model parameters play a crucial role in shaping the dynamics of pressure and energy density within the framework. By adjusting these parameters, we can investigate the behavior of the dynamical aspects of the model more thoroughly. Additionally, the equation of state (EoS) parameter provides a means to probe the late-time acceleration problem. This parameter can be determined utilizing Eqs.~\eqref{first_field_equation} and \eqref{second_field_equation}
        \begin{eqnarray} 
            \omega_{\rm eff} = \frac{p_{\rm eff}}{\rho_{\rm eff}} \label{eq: eos} \, .
        \end{eqnarray}

    To evaluate the theoretical Hubble rate within the $f(R, \mathcal{G})$ framework, the governing Eq.~\eqref{density field eq} is solved numerically. Assuming that matter behaves as a pressureless perfect fluid, the matter density takes the form $\rho_{\mathrm{m}} = 3 H_0^2 \Omega_{\mathrm{m}0} (1+z)^3$, where $z$ is the cosmological redshift defined by $\frac{a_0}{a} = 1+z$. Here $a_0$ is the present-day scale factor, $a$ the scale factor at emission, and $\Omega_{\mathrm{m}0}$ the current matter density parameter. For the specific $f(R, \mathcal{G})$ model under consideration, the first Friedmann equation becomes
        \begin{eqnarray} \label{Eq: ode}
            \hspace{-0.85cm} & & 3 H(z)^2 \Big(6 \big(\alpha  (z+1)^2 H'(z)^2+80 \beta  (z+1)^2 H(z)^4 \nonumber \\ \hspace{-8cm} & & \times H'(z)^2 +2 \alpha  (z+1) H(z) \left((z+1) H''(z)-H'(z)\right) \nonumber\\ \hspace{-8cm} & & +32 \beta  (z+1) H(z)^5 \left((z+1) H''(z)-5 H'(z)\right) \nonumber\\ \hspace{-8cm} & & - 4 \alpha  H(z)^2 - 64 \beta  H(z)^6\big)+1\Big) \nonumber \\ \hspace{-8cm} & & = - 3 H_0^2 \Omega_{\mathrm{m}0} (1+z)^3 
        \end{eqnarray}
    where the prime $(')$ denotes differentiation with respect to $z$.

    Eq. \eqref{Eq: ode} is a second-order differential equation for $H(z)$, requiring two initial conditions for its solution. The first is $H(0) = H_0$, which sets the present-day value of the Hubble parameter. The second can be obtained by ensuring that $H'(0)$ matches the derivative predicted by the standard $\Lambda$CDM expansion law,
        \begin{eqnarray}
            H_{\Lambda \mathrm{CDM}} = H_0 \sqrt{1 - \Omega_{\mathrm{m}0} + \Omega_{\mathrm{m}0} (1+z)^3}.
        \end{eqnarray}
    Differentiating this expression with respect to $z$ and evaluating at $z=0$ yields $H'(0) = \frac{3}{2} H_0 \Omega_{\mathrm{m}0}$. These initial conditions allow the numerical integration of \eqref{Eq: ode}, providing the evolution of the Hubble parameter for the $f(R, \mathcal{G})$ scenario.

\section{Fitting \texorpdfstring{$f(R, \mathcal{G})$}{} Gravity Models to \texorpdfstring{$\Lambda$}{}CDM using Gradient Descent} \label{SEC-3}
    Gradient descent is widely used in cosmology as an optimization algorithm, primarily within machine learning applications to analyze observational data, optimize model parameters, and refine simulations. Standard cosmological parameter estimation typically relies on Bayesian methods like Markov Chain Monte Carlo (MCMC) sampling. However, when fitting complex modified gravity models to large datasets, gradient descent offers computational advantages \cite{Krippendorf_2022_3, Olvera_2022_8, Anandam_2023_100}. Instead of randomly sampling parameter space, this method systematically moves toward the best fit by following the steepest decrease in the cost function.    

    When testing such models against the concordance $\Lambda$CDM background, it is useful to match the predicted expansion history to the reference curve over a given range of redshift. This is where Gradient Descent can be a practical fitting tool. The method starts by defining a cost function that measures the difference between the model predictions and the $\Lambda$CDM quantities of interest, such as the Hubble rate or the deceleration parameter. If the set $\{z_i\}_{i=1}^{N}$ represents the sampling points in redshift space, with $H_{f(R, \mathcal{G})}(z_i; \theta)$ the value predicted by the $f(R, \mathcal{G})$ model and $H_{\Lambda \mathrm{CDM}}(z_i)$ the corresponding $\Lambda$CDM value. For our $f(R,\mathcal{G})$ model, we define the cost function as the mean squared difference between predicted and observed Hubble rates
        \begin{eqnarray}
            J(\theta) = \frac{1}{N} \sum_{i=1}^{N} \left[ H_{f(R, \mathcal{G})}(z_i; \theta) - H_{\Lambda \mathrm{CDM}}(z_i) \right]^2 ,
        \end{eqnarray}
    where $N$ is the total number of redshift samples used in the comparison. Dividing by $N$ normalizes the cost function, making its value independent of the number of samples and ensuring that changes in $N$ do not trivially scale the optimization step sizes. The algorithm updates parameters iteratively: if the current prediction overshoots the data, gradients tell us exactly how to adjust $\alpha$ and $\beta$ to reduce this error.

    The Gradient Descent algorithm updates the parameter vector $\theta$ according to
        \begin{eqnarray}
            \theta^{(k+1)} = \theta^{(k)} - \eta \, \nabla_{\theta} J(\theta^{(k)}) ,
        \end{eqnarray}
    where $k$ denotes the iteration step and $\eta$ is a small positive number controlling how far each update moves in parameter space. The gradient of the cost function with respect to each parameter involves the residuals between model and reference values multiplied by the partial derivatives of the model predictions. If these derivatives are not available analytically, they can be estimated through finite differences, though automatic differentiation offers greater accuracy when feasible.

    Starting from an initial guess, the parameters are repeatedly adjusted to reduce $J(\theta)$. A careful choice of learning rate is important. Too small a value slows convergence, while a value that is too large can cause the parameters to overshoot the minimum and oscillate without settling. The number of redshift points $N$ also influences the fit. If $N$ is small, the optimization may not capture the shape of the $\Lambda$CDM curve across the full range, while a very large $N$ increases computational cost and can require tuning of $\eta$ to maintain stability.

    Our numerical $f(R, \mathcal{G})$ model emerges from solving Eq.~\eqref{Eq: ode} through the \texttt{ODEint} method, with a set of carefully chosen initial conditions. Once the numerical solution is obtained, we compare the model output with the $\Lambda$CDM predictions. The comparison is carried out using the Gradient Descent technique, a numerical optimization approach that iteratively adjusts the free parameters to minimize the difference between the two curves. This is achieved by calculating the gradient of the error function with respect to the parameters and updating them in the opposite direction of the gradient, step by step, until convergence. The method is particularly effective for smooth and differentiable models, enabling us to identify the optimal set of parameters for the $f(R, \mathcal{G})$ case. In the context of $f(R, \mathcal{G})$ gravity, this fitting process allows the expansion history of the modified model to be tuned so that it tracks the standard cosmology to high accuracy. Applying this method to our specific functional form of $f(R, \mathcal{G})$, we determined the best fit parameter values to be $\alpha=0.562 \pm 0.045$, $\beta=1.3 \pm 0.141$, $\Omega_\mathrm{m0}=0.3 \pm 0.002$ and $H_0=70.102 \pm 1.214\, \mathrm{km}\, \mathrm{s}^{-1} \, \mathrm{Mpc}^{-1}$ using the Gradient Descent technique for 32 CC samples and 1701 Pantheon$^+$ supernovae samples. All graphs in this study are based on the combined analysis of the CC and Pantheon$^+$ datasets, offering a comprehensive assessment of model performance across diverse observational data. In the following section, we examine the cosmological parameters, energy conditions, $\mathrm{Om}(z)$ diagnostics, and the estimated age of the Universe.    

\subsection{Equations of State and Deceleration Parameter}
\begin{figure}[!htp]
    \centering
        \includegraphics[scale=0.75]{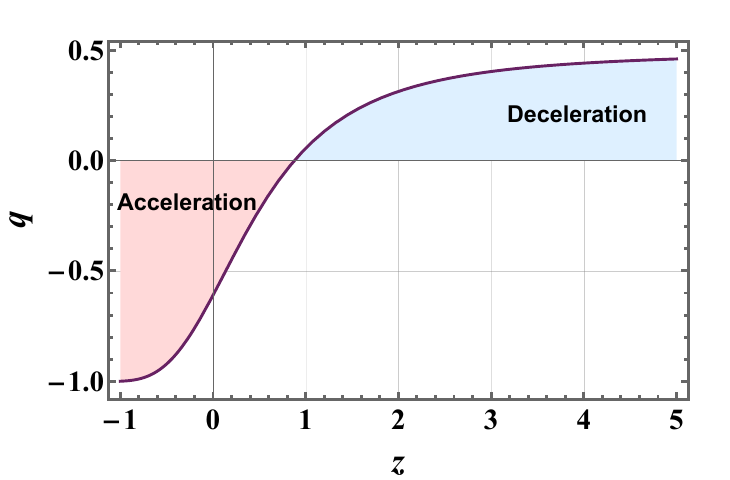}
        \caption{Evolution of the deceleration parameter $q(z)$ from past deceleration to current acceleration, with transition at $z_{\mathrm{t}}=0.763$ based on combined CC+Pantheon$^+$ datasets.}
    \label{Fig: q}
    \end{figure}

\begin{figure}[!htp]
    \centering
        \includegraphics[scale=0.75]{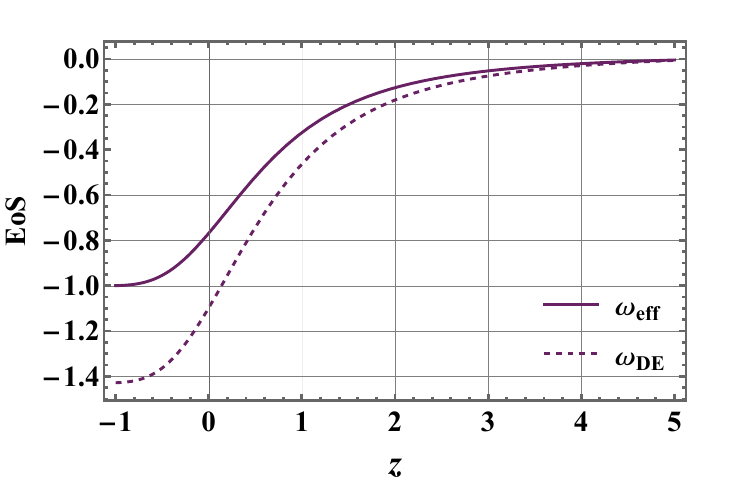}
        \caption{Evolution of the effective and dark energy EoS parameter $\omega_{\rm eff}$ with redshift, reconstructed from the combined CC and Pantheon$^+$ datasets.}
    \label{Fig: EoS}
    \end{figure}

    We analyze the evolution of the deceleration parameter and the EoS in this section, providing insights into the expansion history of the Universe and the dynamical properties of cosmic components. Fig.~\ref{Fig: q} shows the deceleration parameter $q(z)$ as reconstructed from CC and Pantheon$^+$ supernovae data measurements.  A positive $q$ value corresponds to a slowing expansion, while a negative value signals acceleration.  The curves cross zero at redshifts $z_{\mathrm{t}}=0.763$ for the CC+Pantheon$^+$ dataset combinations, closely matching transition redshifts reported by independent analyses.  Present day values of $q_{0}=-0.612$ lie within the range $q_{0}=-0.528_{-0.088}^{+0.092}$ from recent observational studies \cite{Gruber_2014_89}. Fig.~\ref{Fig: EoS} shows the quintessence-like dynamics approaching the cosmological constant limit of $-1$ in the late-time universe, indicating convergence towards the $\Lambda$CDM model. The current EoS at $z = 0$ is computed to be $-0.785$, derived from the analysis of combined datasets. From the evolution of the effective equation-of-state (EoS) parameter and its connection to the dark-energy EoS, we find that the present value $\omega_{\rm eff} = -0.785$ and $\omega_{\rm DE}=-1.098$ is consistent with observational constraints from Planck 2018 $\omega_{\text{DE}}=-1.03\pm 0.03$ \cite{Aghanim_2020_641}, and WAMP+CMB $\omega_{\text{DE}}=-1.079^{+0.090}_{-0.089}$ \cite{Hinshaw_2013_208}. The smooth evolution of both parameters indicates that the $f(R,\mathcal{G})$ model naturally reproduces quintessence-like behavior through geometric modifications alone.

\section{Energy Conditions} \label{energy conditions}
    In the context of GR, energy conditions have long served as powerful tools for drawing broad and general conclusions about the behavior of strong gravitational fields and the structure of cosmological spacetime \cite{Hawking_1973_book, Poisson_2004_book}. They establish a connection between the structure of spacetime and the characteristics of matter by placing restrictions on the stress–energy tensor that ensure a physically sensible distribution of energy. Within classical GR, these conditions are often applied to examine singularity formation, the evolution of null, spacelike, and timelike geodesics, and the causal structure of cosmological models \cite{Santos_2007_76, Kar_2007_69}.

    A particularly useful geometric starting point for defining these conditions is the Raychaudhuri equation, which describes the evolution of a congruence of geodesics without reference to the specific gravitational field equations. For a null congruence with vanishing vorticity, the equation takes the form
        \begin{equation} \label{eq.19}
            \frac{d\theta}{d\tau}=-\frac{1}{2} \theta^2-\sigma_{ab} \sigma^{ab}-R_{ab} k^a k^b,
        \end{equation}
    where $\theta$ is the expansion scalar, $\sigma_{ab}$ is the shear tensor, and $k^a$ is a null vector. Since the shear term is non–negative, attractive gravity requires $R_{ab}k^a k^b \geq 0$. In the framework of Einstein’s equations, this becomes the NEC condition $\mathcal{T}_{ab}k^a k^b \geq 0$. Similarly, the WEC follows from $\mathcal{T}_{ab}u^a u^b \geq 0$ for a timelike vector $u^a$.

    The usual classification includes the null, weak, strong, and dominant energy conditions, each reflecting a different physical requirement. The null energy condition (NEC) argues that for any null geodesic, the combined value of the energy density $\rho$ and the pressure $p$ must be at least zero, which can be expressed mathematically as $\rho + p \geq 0$. This principle is critical within the realm of GR, particularly regarding the stability of spacetime and the emergence of singularities. The weak energy condition (WEC) requires that the local energy density is non-negative for every timelike observer, resulting in the constraints $\rho \geq 0$ and $\rho + p \geq 0$. This indicates that both the energy density $\rho$ and the total of energy density plus pressure must be at least zero. The strong energy condition (SEC) encodes the expectation that gravity remains attractive, which in a perfect fluid form reads $\rho + 3p \geq 0$ together with $\rho + p \geq 0$. The dominant energy condition (DEC) further ensures that energy flow is causal, with $\rho \geq 0$ and $\rho \pm p \geq 0$. A violation of the NEC automatically signals the breakdown of all other standard energy conditions.

\begin{figure}[!htp]
    \centering
        \includegraphics[scale=0.75]{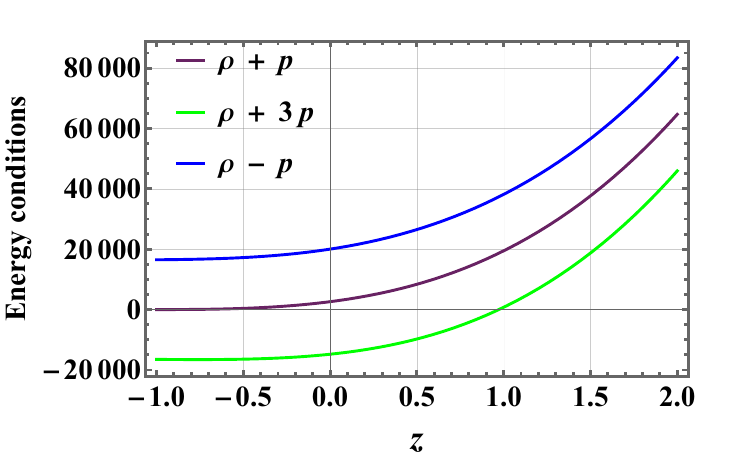}
        \caption{Graphical representation of the evolution of energy conditions as a function of redshift $z$. The plot illustrates how the different energy conditions behave across cosmic history.}
    \label{Fig: EC}
    \end{figure}

    In cosmology, the SEC has drawn particular attention because it must be violated during inflation and again in the present era to account for the observed accelerated expansion \cite{Barcelo_2002_11, Visser_1997_56}. In Fig.~\ref{Fig: EC}, our examination of the $f(R, \mathcal{G})$ model reveals that the NEC and the DEC are upheld for the majority of cosmic evolution. However, it is noteworthy that the SEC is violated in the late-time regime. This violation emerges around $z \approx 0.97$ for the datasets considered, marking the onset of accelerated expansion. The WEC remains positive from the early epoch to the present, consistent with a quintessential form of dark energy. The detailed redshift evolution of these conditions is presented in Fig.~\ref{Fig: EC}, showing that the transition in SEC behavior aligns closely with the change in cosmic acceleration. In Fig. \ref{Fig: EC}, we utilized the parameter values determined through Gradient Descent optimization. Notably, varying the parameters $\alpha$ and $\beta$ within their 1$\sigma$ confidence intervals does not result in any qualitative changes to the energy-condition curves. A complete understanding of these results benefits from fixing the Hubble rate through observational constraints or theoretical assumptions, which in turn allows for a more precise mapping of the energy condition boundaries within the $f(R, \mathcal{G})$ scenario.

\section{Age of the Universe and \texorpdfstring{$\mathrm{Om}(z)$}{} diagnostic} \label{SEC-4}
    In our $f(R,\mathcal{G})$ gravity scenario, we first examine the predicted age of the Universe using the CC+Pantheon$^+$ supernovae dataset as the observational reference. The cosmic age at a given redshift $z$ can be expressed in terms of the Hubble expansion rate $H(z)$, where the variable of integration is the redshift $x$. Using this relation, the difference in cosmic time between redshift $z$ and the present epoch is obtained by integrating the same expression from today up to that redshift   
    \begin{eqnarray}
        H_0 (t_0 - t) = \int_{0}^{z} \frac{dx}{(1+x) E(x)}, \quad E(z) = \frac{H^2(z)}{H_0^2} ,
    \end{eqnarray}
    where $H_0$ is the present-day Hubble constant. Taking the limit $z \to \infty$ yields the total age of the Universe,
    \begin{eqnarray}
        H_0 t_0 = \lim_{z \to \infty} \int_{0}^{z} \frac{dx}{(1+x) E(x)} .
    \end{eqnarray}
    From this relation, the quantity $1/H_0$ serves as a natural scale for cosmic age, typically modified only by a factor very close to unity.
    \begin{figure}[!htp]
    \centering
        \includegraphics[scale=0.75]{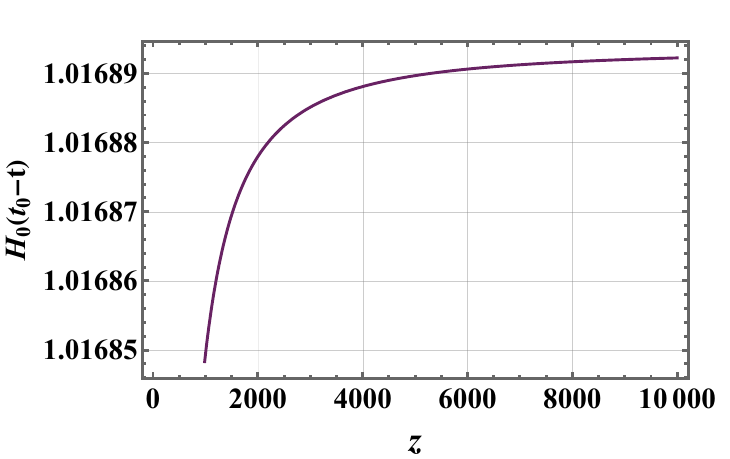}
        \caption{Graphical representation of cosmic time as a function of redshift.}
    \label{Fig: AoU}
    \end{figure}

    Using the best-fit parameters of our numerical $f(R,\mathcal{G})$ model—obtained from solving Eq.~\eqref{Eq: ode} with chosen initial conditions and matched to $\Lambda$CDM through a Gradient Descent procedure—we find for Pantheon data that $H_0(t_0 - t)$ approaches $1.01689$ as $z \to \infty$ as shown in Fig.~\ref{Fig: AoU}. This corresponds to a present cosmic age of $t_0 \approx 14.011$ Gyr, which is in close agreement with $13.8 \pm 4$ \cite{Cowan_2002_572} and Planck measurements of $t_0 = 13.786 \pm 0.020$ Gyr \cite{Aghanim_2020_641}. This agreement suggests that the model reproduces the expected timeline of cosmic history and remains compatible with the ages of the oldest observed stars. Our best-fit cosmic age is 14.011 Gyr, compared to the Planck value of $13.786 \pm 0.020$ Gyr, yielding a modest offset of $\sim 0.225$ Gyr (about $1.6\%$). This discrepancy is primarily driven by the higher best-fit $H_0$ obtained in our analysis and by methodological differences, as our constraints are derived from late-time observables rather than the Planck CMB likelihood.

    After establishing the age consistency, we turn to the $\mathrm{Om}(z)$ diagnostic, a purely geometric probe designed to distinguish dark energy models without relying on an assumed matter density. For a given $z$, it is defined as
    \begin{eqnarray}
        \mathrm{Om}(z) = \frac{E(z) - 1}{(1+z)^3 - 1} ,
    \end{eqnarray}
    with $E(z) = H^2(z) / H_0^2$. In the two-point version,
    \begin{eqnarray}
        \mathrm{Om}(z_1, z_2) = \mathrm{Om}(z_1) - \mathrm{Om}(z_2),
    \end{eqnarray}
    positive values ($z_1 < z_2$) indicate quintessence-like behavior ($\omega > -1$), negative values indicate phantom behavior ($\omega < -1$), and a constant $\mathrm{Om}(z)$ across redshift marks the $\Lambda$CDM case. The slope of $\mathrm{Om}(z)$ thus acts as a quick identifier for the nature of dark energy.
    \begin{figure}[!htp]
    \centering
        \includegraphics[scale=0.75]{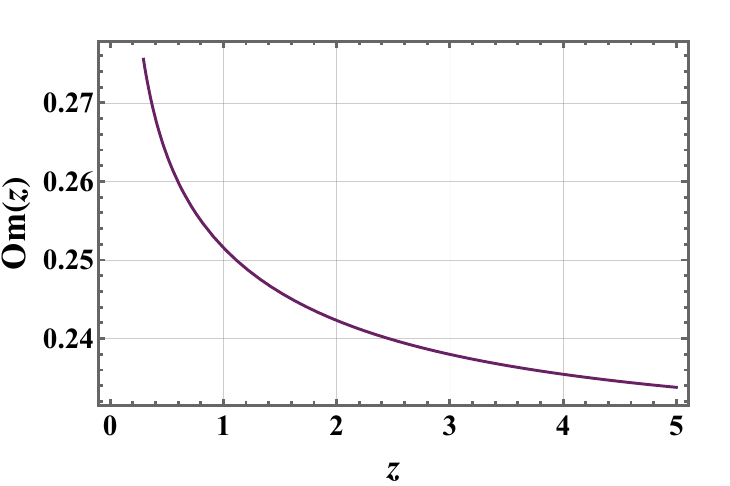}
        \caption{Graphical representation of the evolution of $\mathrm{Om}(z)$ as a function of redshift $z$.}
    \label{Fig: Omz}
    \end{figure}
    
    In Fig.~\ref{Fig: Omz}, our reconstructed $\mathrm{Om}(z)$ profile exhibits a decreasing trend with increasing redshift, indicating a quintessence-like evolution of dark energy within the fitted $f(R,\mathcal{G})$ framework for the combined CC+Pantheon$^+$ dataset. This pattern reinforces the view that the model can mimic $\Lambda$CDM at the background level while allowing subtle deviations that might be testable with more precise data.

\section{Conclusion} \label{SEC-5}
    The application of Gradient Descent optimization techniques has proven essential for constraining cosmological parameters in modified gravity frameworks, particularly in $f(R,\mathcal{G})$ gravity. These iterative algorithms minimize chi-squared functions by computing parameter gradients and updating values in the steepest descent direction, offering computational advantages over traditional MCMC methods. Modern implementations incorporate adaptive learning rates and momentum terms to enhance convergence stability, enabling efficient exploration of high-dimensional parameter spaces characteristic of $f(R, \mathcal{G})$ theories. The Gradient Descent approach proves especially valuable when analyzing observational datasets like Pantheon supernovae, where complex modified gravity equations require robust optimization strategies to extract reliable cosmological constraints while maintaining computational tractability across extensive redshift ranges.

    Our comprehensive investigation of $f(R, \mathcal{G})$ gravity using CC+Pantheon$^+$ observational data establishes this theoretical framework as a compelling alternative to standard dark energy cosmology through geometric modifications of spacetime curvature. The derived cosmic age of approximately 14.011 Gyr demonstrates remarkable consistency with with $13.8 \pm 4$ \cite{Cowan_2002_572} and Planck satellite measurements of $13.786 \pm 0.020$ Gyr, validating the modified gravity approach while maintaining compatibility with independent age determinations. The Om diagnostic analysis reveals quintessence-like behavior throughout cosmic evolution, with values decreasing from early to late times and indicating effective dark energy behavior characterized by EoS parameter $\omega > -1$, distinguishing our model from phantom energy scenarios. Energy condition investigations confirm that NEC and DEC remain satisfied while SEC violation occurs at transition redshifts $z \approx 0.97$, precisely when cosmic acceleration commences and provides the necessary mechanism for observed late-time expansion. The deceleration parameter exhibits a smooth evolution from positive to negative values with the present value of the EoS parameter $\omega_{\rm eff} \approx -0.785$ and $\omega_{\rm DE} \approx -1.098$ placing the model within the quintessence regime, collectively establishing $f(R,\mathcal{G})$ gravity as a theoretically consistent and observationally viable explanation for cosmic acceleration through fundamental gravitational modifications rather than components of exotic matter.

\section*{Acknowledgments} BM acknowledges the support of Council of Scientific and Industrial Research (CSIR) for the project grant (No. 03/1493/23/EMR II).

\providecommand{\href}[2]{#2}\begingroup\raggedright\endgroup

\end{document}

%% file: preamble.tex
\usepackage{amsfonts,amssymb}
\usepackage{graphicx}

\def\noi{\noindent}

\newcommand{\Title}[1]{\noi {{\Large\bf #1}}\\[1ex]}

\def\Aunames#1{\noi{\bf #1}}
\def\au#1{${}^{#1}$}
\def\Addresses#1{\medskip\noi \protect
	\begin{description}\itemsep -3pt {\it #1} \end{description}}
\def\adr#1#2{\item[${}^{#1}$]{\it #2}}

\newcommand{\Abstract}[1]{\vskip 2mm \begin{center}
        \parbox{16.4cm}{\small\noi #1} \end{center}\medskip}

\def\email#1#2{\footnotetext[#1]{e-mail: #2}\addtocounter{footnote}{1}}

\def\nqq{\hspace*{-2em}}

\usepackage{color}
\def\red#1{{\color{red} #1}}

\def\Jl#1#2{#1 {\bf #2},\ }

\def\ApJ#1 {\Jl{Astroph. J.}{#1}}
\def\CQG#1 {\Jl{Class. Quantum Grav.}{#1}}
\def\DAN#1 {\Jl{Dokl. AN SSSR}{#1}}
\def\GC#1 {\Jl{Grav. Cosmol.}{#1}}
\def\GRG#1 {\Jl{Gen. Rel. Grav.}{#1}}
\def\IJMPD#1 {\Jl{Int. J. Mod. Phys. D}{#1}}
\def\JETF#1 {\Jl{Zh. Eksp. Teor. Fiz.}{#1}}
\def\JETP#1 {\Jl{Sov. Phys. JETP}{#1}}
\def\JHEP#1 {\Jl{JHEP}{#1}}
\def\JMP#1 {\Jl{J. Math. Phys.}{#1}}
\def\NPB#1 {\Jl{Nucl. Phys. B}{#1}}
\def\NP#1 {\Jl{Nucl. Phys.}{#1}}
\def\PLA#1 {\Jl{Phys. Lett. A}{#1}}
\def\PLB#1 {\Jl{Phys. Lett. B}{#1}}
\def\PRD#1 {\Jl{Phys. Rev. D}{#1}}
\def\PRL#1 {\Jl{Phys. Rev. Lett.}{#1}}

\def\lal{&&\nqq {}}

\def\beq{\begin{equation}}
\def\eeq{\end{equation}}
\def\bear{\begin{eqnarray}}
\def\bearr{\begin{eqnarray} \lal}
\def\ear{\end{eqnarray}}
\def\earn{\nonumber \end{eqnarray}}

